\documentclass[journal]{IEEEtran}
\usepackage{mathrsfs}
\usepackage{multirow}
\usepackage{booktabs}
\usepackage{hyperref}
\usepackage{color}
\usepackage{amsmath, bm}
\usepackage[ruled,linesnumbered]{algorithm2e}
\usepackage{tabu}

\usepackage{flushend}

\usepackage{mathrsfs}
\ifCLASSINFOpdf
  \usepackage[pdftex]{graphicx}
  \graphicspath{{../pdf/}{../jpeg/}{./figure/photo/}}
\else
  \usepackage[dvips]{graphicx}
  \graphicspath{{../eps/}}
\fi

\usepackage{amsmath,amsthm,amssymb}

\begin{document}
%
\title{BaMBNet: A Blur-aware Multi-branch Network for Defocus Deblurring}
%
%
%

\author{Pengwei Liang,
        Junjun~Jiang,~\IEEEmembership{Member,~IEEE,~}
        Xianming Liu,~\IEEEmembership{Member,~IEEE,~}
        and Jiayi Ma,~\IEEEmembership{Member,~IEEE}

\thanks{The research was supported by the National Natural Science Foundation of China (61971165, 61922027, 61773295), and also is supported by the Fundamental Research Funds for the Central Universities. } %
\IEEEcompsocitemizethanks{
\IEEEcompsocthanksitem P. Liang, J. Jiang, and X. Liu are with the School of Computer Science and Technology, Harbin Institute of Technology, Harbin 150001, China. E-mail: \{jiangjunjun, csxm\}@hit.edu.cn.
\IEEEcompsocthanksitem J. Ma is with the Electronic Information School, Wuhan University, Wuhan 430072, China. E-mail: jyma2010@gmail.com.
}
}

\markboth{ }%
{Shell \MakeLowercase{\textit{\emph{et al.}}}: Bare Demo of IEEEtran.cls for Journals}
%



\maketitle

\begin{abstract}
The defocus deblurring raised from the finite aperture size and exposure time is an essential problem in the computational photography. It is very challenging because the blur kernel is spatially varying and difficult to estimate by traditional methods. Due to its great breakthrough in low-level tasks, convolutional neural networks (CNNs) have been introduced to the defocus deblurring problem and achieved significant progress. However, they apply the same kernel for different regions of the defocus blurred images, thus it is difficult to handle these nonuniform blurred images. To this end, this study designs a novel blur-aware multi-branch network (BaMBNet), in which different regions (with different blur amounts) should be treated differentially. In particular, we estimate the blur amounts of different regions by the internal geometric constraint of the DP data, which measures the defocus disparity between the left and right views. Based on the assumption that different image regions with different blur amounts have different deblurring difficulties, we leverage different networks with different capacities (\emph{i.e.} parameters) to process different image regions. Moreover, we introduce a meta-learning defocus mask generation algorithm to assign each pixel to a proper branch. In this way, we can expect to well maintain the information of the clear regions while recovering the missing details of the blurred regions. Both quantitative and qualitative experiments demonstrate that our BaMBNet outperforms the state-of-the-art methods. Source code will be available at \url{https://github.com/junjun-jiang/BaMBNet}.
\end{abstract}


\section{Introduction}
\label{sec:intro}
Defocus blurring is inevitable when the scene regions (with wider depth range) are out-of-focus due to the limitation of the hardware, \emph{i.e.}, cameras with a finite-size aperture can only focus on the shadow depth of field (DoF) at a time, and the rest scene regions will contain blur \cite{abuolaim2020defocus}. Removing this blur and recovering defocused image details are challenging due to the spatially-varying point spread functions (PSFs) \cite{veeraraghavan2007dappled, chaudhuri2012depth, xu2020towards}. Recently, some studies address this problem by dual-pixel (DP) sensors found on most
modern cameras \cite{abuolaim2020learning}. Although the DP sensors were originally designed to facilitate autofocus \cite{sliwinski2013simple,jang2015sensor,herrmann2020learning}, it has been found to be very useful in a wide range of applications such as depth
estimation \cite{garg2019learning}, defocus deblurring \cite{pan2020dual}, reflection removal \cite{punnappurath2019reflection}, and synthetic DoF \cite{zhang20202}. DP sensors provide a pair of photodiodes for each pixel location to capture two sub-aperture views of the same scene \cite{wadhwa2018synthetic,vadathya2019unified}. Compared with a single photodiode for each pixel in the traditional sensor, the two sub-aperture blurred views provide more information for spatially-varying blur detection and defocus deblurring \cite{punnappurath2020modeling}.

\begin{figure}
    \centering
    \includegraphics[width=0.9\linewidth]{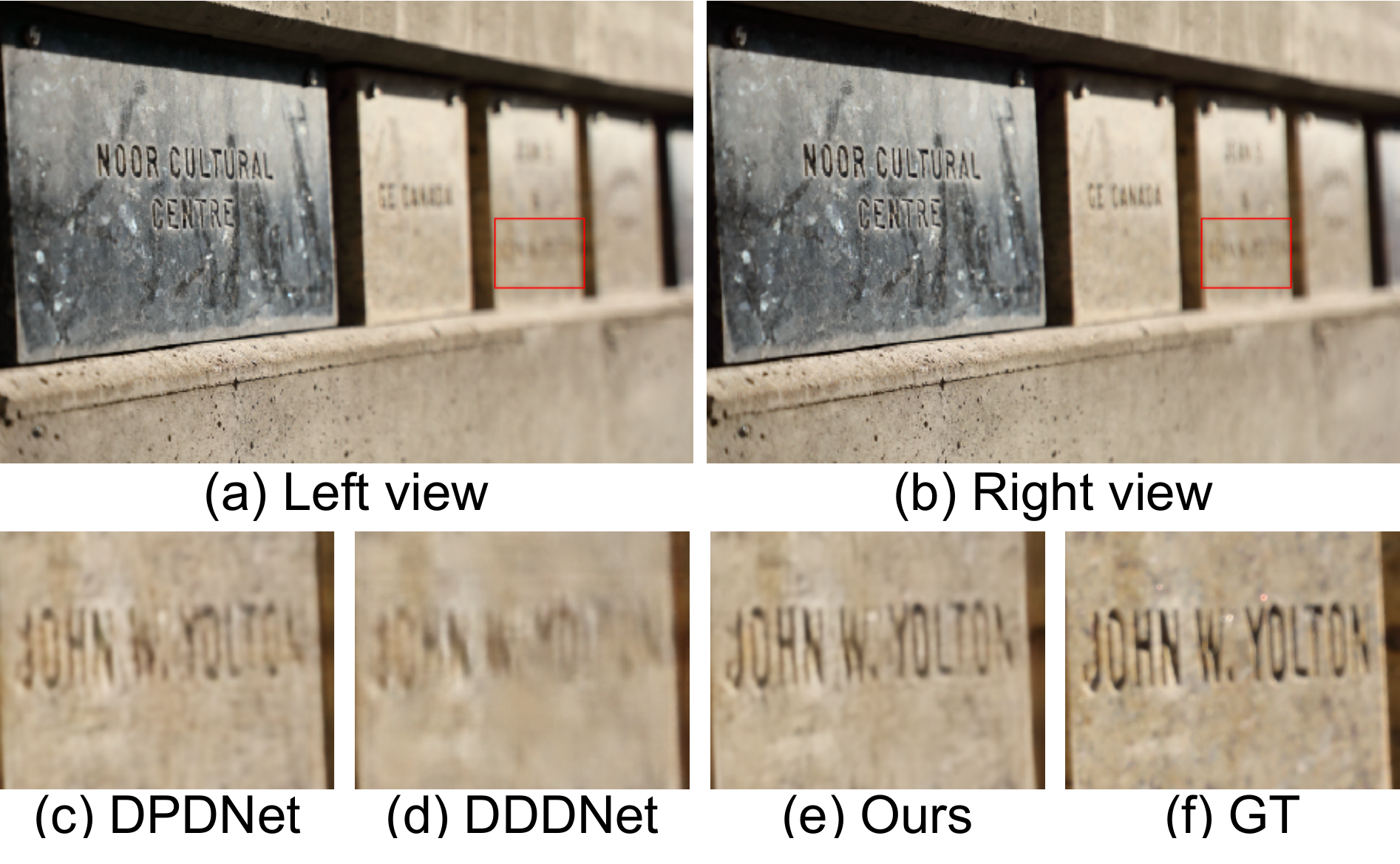}
    \caption{Schematic illustration of defocus deblurring. (a) and (b) shows a pair of blurred input images. For convenience, we will only show the left view as the input image in the following figures. (c-f) show the highlighted deblurred results of DDDNet \cite{pan2020dual}, DPDNet \cite{abuolaim2020defocus} , and BaMBNet (Ours), respectively. }
    \label{fig:indtroduction}
\end{figure}

As shown in Fig. \ref{fig:indtroduction}(a), the blur image can be approximately divided into two categories, in-focus and out-of-focus, which correspond to the sharp regions and blurred regions, respectively. We expect that the deblurred results can keep the details of in-focus regions while sharpening the blurred regions. Thanks to the immense success of deep learning, in the most recent years some end-to-end deep neural networks such as DPDNET \cite{abuolaim2020defocus} and DDDNet \cite{pan2020dual} have achieved pleasing deblurring results. However, they tackle sharp regions and blurred regions by the same deep convolution network, and it is a great challenge for a single network to balance between keeping the details in the focus regions and deblurring in the out-of-focus regions. For instance, the highlighted patches in Fig. \ref{fig:indtroduction}(c-d) indicate that methods based on the single network may fail to handle scenes with larger depth variation.

According to \cite{hecht2001optics}, the blur amounts that measure the blur levels of one image are varying with respect to different regisons and can be determined via the circle of confusion (COC) size $c(d)$:
\begin{equation}
    c(d)  = \frac{| d - d_f|}{d} \cdot \frac{f_0^2}{N( d_f - f_0)},
\label{Imaging}
\end{equation}
where $d$ is the subject-to-camera distance, $d_f$ is the focus distance,  $f_0$ and $N$ are the focal length and the stop number of the camera, respectively. The $f_0$, $N$ and $d_f$ are related to the camera settings and are usually fixed before shooting.
From the definition of COC, we can learn that the blur amounts change with the depth. However, previous end-to-end deep network methods \cite{abuolaim2020defocus, pan2020dual} ignore this observation. In the balance between deblurring and maintaining clear regions, they tend to fall into a ``trivial solution", where gains in one thing and losses in another. This observation motivates us to carefully exploit more efficient deblurring methods according to the blur amounts of different regions (depths). In other words, how to accurately estimate the blur amounts of image, \emph{i.e.} the COC map, and effectively integrate it into the deblurring procedure is the critical problem.


To combat these challenges, we propose a blur-aware multi-branch network to address the defocus deblurring problem. 
In practice, we first estimate the COC map of the input image pair and then transform the COC map into defocus masks by a meta-learning mechanism, which can assign different image pixels to different branch networks.
Based on the assumption that recovering the blurred regions requires considerable learning parameters while maintaining clear regions only needs a few parameters, we apply different branch networks to deal with different regions under the guidance of defocus masks. In this way, the lightest branch with the fewest learning parameters will only pay attention to the in-focus regions and maintain the clear regions of the input images. In contrast, the heaviest branch with the most parameters is used to reconstruct the missing details and recover the sharp parts from the regions with a large amount of blur. In other words, if the branch has more parameters, we expect it to focus on a more blurred region. In addition, compared with estimating a depth map to implicitly guide the defocus deblurring \cite{pan2020dual}, our COC map is estimated in an unsupervised way and does not require an additional ground truth. Moreover, since the defocus masks are used to guide to assign the regions into proper branches in our method, the entire procedure can be regarded as a divide-and-conquer strategy. The main idea of the strategy is decomposing the source problem into multiple easy sub-problems, and our model is prone to optimize with the assistance of defocus masks.
We carry out comparison experiments between the proposed method and traditional hand-crafted one in defocus deblurring.
The proposed approach is also compared with existing deep defocus deblurring approaches to demonstrate the effectiveness of blur-aware multi-branch network.

The contributions of this work can be summarized as follows:
\begin{itemize}
    \item We propose a blur-aware multi-branch network (BaMBNet) to address the problem of non-uniform blur distribution in realistic defocus images. Different regions with different blur amounts will be treated by multiple branches with different capacities, therefore, our method can well maintain the information of the clear regions while recovering the missing details of the blurred regions.
    \item We propose a COC map estimation method in an unsupervised way, thus avoiding the requirement for the ground truth. To effective guide the optimization of multi-branch network, with the estimated COC map we introduce a meta-learning strategy to generate the defocus masks.
    \item Extensive experiments demonstrate that our method outperforms the existing state-of-the-art approaches(SOTAs) such as DPDNET \cite{abuolaim2020defocus} and DDDNet \cite{pan2020dual}. The ablation studies also verify the effectiveness of different components of the proposed method.
\end{itemize}

The remainder of this paper is organized as follows.
Section \ref{sec:related} introduces the dual pixel and reviews existing defocus deblurring methods in the
literature. Section \ref{sec:Proposed} presents our image defocus deblurring network
and the proposed COC estimation and assignment strateyies. Section \ref{sec:Experiments} provides the comparison experiments with SOTAs and demonstrates the technical contributions of the proposed method in the ablation studies. Section \ref{sec:Conclusions} concludes this paper.

\section{Related Work}
\label{sec:related}
In recent years, dual pixel has come into fashion in low-level vision tasks such as depth estimation and defocus deblurring \cite{garg2019learning, pan2020dual, abuolaim2020defocus}. In this section, we will introduce the dual pixel in brief and summarize various defocus deblurring methods.

\subsection{Dual-Pixel Camera Model}
\begin{figure}[htbp]
    \centering
    \includegraphics[width=0.9\linewidth]{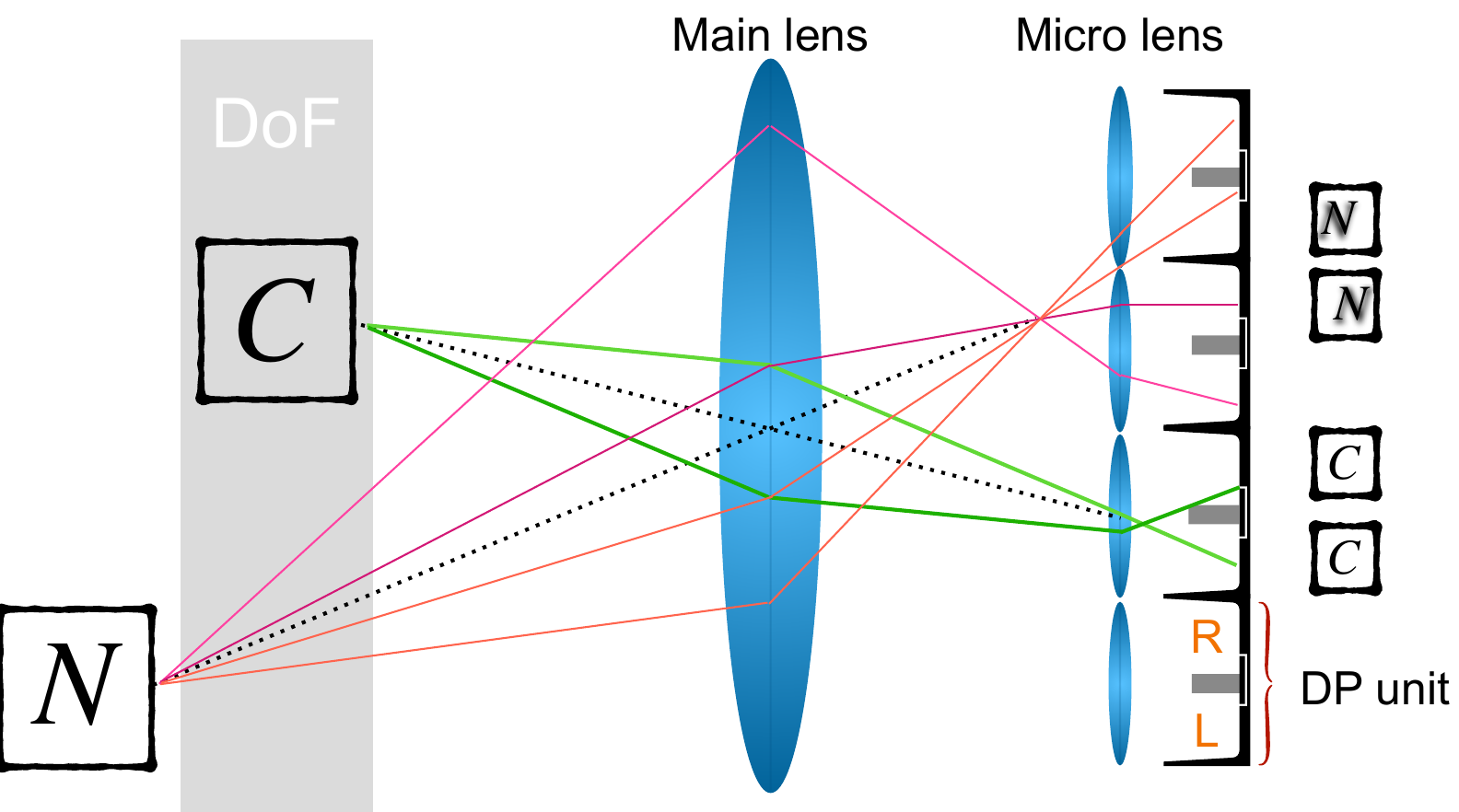}
    \caption{Optical geometry of dual-pixel camera based on thin-lens model.}
    \label{fig:dual-pixel}
\end{figure}

A DP sensor allocates a microlens and a pair of photodiodes for each pixel, as shown in Fig. \ref{fig:dual-pixel}. Each photodiode can record the light ray independently. In other words, there will be two views of the same scene captured by a DP camera, called the left view and the right view \cite{abuolaim2020learning,punnappurath2019reflection,wadhwa2018synthetic}.
When the region is far away from the focus plane, there will be detectable disparity in the left and right views, which is referred to as \textit{defocus disparity}. By measuring the level of \textit{defocus disparity}, the autofocus routine can adjust the lens movement to bring the out-of-focus regions into focus, which has been widely applied in business. Recently, some studies show that the \textit{defocus disparity} also can be used for depth estimation, reflection removal, defocus deblurring, \emph{etc.} Next, we will describe the defocus disparity in detail with two representative examples.

Fig. \ref{fig:dual-pixel} illustrates an interesting phenomenon. As we can see, there is an object recorded by the DP camera located in the DoF region, \emph{i.e.} the character `C'. In this case, the light rays striking from different angles are projected into the surface of the micro lens, and each photodiode registers the average of all the light rays.
As a result, the formed left and right views are very close, \emph{i.e.} the character `C' is clear, which is the same as the imaging results of a traditional non-DP camera. In contrast to the DoF region, if the object is placed far away from the DoF region, \emph{e.g.} the character `N',
the light ray originating from the object will converge at a point away from the plane of the micro lens and create a few pixel wide blur on the sensor, \emph{e.g.} the blur results of character `N' in Fig. \ref{fig:dual-pixel}. Since the two photodiodes record different striking angles, the final left and right blur views show the \textit{defocus disparity}.
Now consider what occurs if we replace the DP camera with a traditional non-DP camera and keep the same settings.
The blurred regions will show more blur amounts with respect to non-DP camera due to all light rays striking the micro lens are gathered into one photodiode \cite{adelson1992single}.

\subsection{Defocus Deblurring}
According to the procedure of deblurring, the technique of defocus deblurring can be summarized into two categories: (i) one is that two-stage cascade approaches consists of two steps, where the first stage is defocus detection, and the second stage is non-blind deblurring by deconvolution \cite{d2016non,karaali2017edge,lee2019deep, park2017unified,shi2015just, tang2019defusionnet}, (ii) the other is the one-stage end-to-end methods \cite{abuolaim2020defocus, pan2020dual}.

In the two-stage methods, a common strategy is to first estimate the defocus map and then use a deconvolution to recover the out-of-focus regions indicated by the defocus map. Defocus map estimation is the more important stage of the two stages. Representative works include Karaali \emph{et al.} \cite{karaali2017edge}, who used identifiable hand-crafted features such as image gradients to calculate the difference between the original image edges and the re-blurred image edges. Besides, other similar methods include Shi \emph{et al.} \cite{shi2014discriminative} using the edge representation and Yi \emph{et al.} \cite{yi2016lbp} using a local binary pattern to measure the focus sharpness. Recently, some studies have used the learning-based end-to-end networks to estimate the defocus map. For example, Park \emph{et al.} \cite{park2017unified} combined the deep features and hand-crafted features together to estimate the blur amounts on edges. Following the Park \emph{et al.} \cite{park2017unified}, Zhao \emph{et al.} \cite{zhao2018defocus,zhao2019defocus} proposed a Fully Convolutional Network which is robust to scale transformation. In addition, Lee \emph{et al.} \cite{lee2019deep} introduced a large-scale dataset for DNN-based training and estimated dense defocus maps via domain adaption. Nevertheless, the common disadvantage of these methods is that the defocus map is asked to convert as a binary mask before second stage, so that the information of the estimated defocus map would not be fully utilized.

In the end-to-end learning-based methods, Abdullah \emph{et al.} \cite{abuolaim2020defocus} firstly introduced DPDNet to address defocus deblurring on DP images, and they simultaneously released a supervised in-the-wild defocus deblurring dataset. The DPDNet achieves better performance when compared with these two-stage deep learning-based methods. However, DPDNet does not explicitly extract the latent blur amounts of DP pairs and treat different regions indiscriminately. Therefore, it may blur the clean region or cannot well recover the blur region. After that, Pan \emph{et al.} \cite{pan2020dual} proposed to jointly perform the defocus deblurring and depth estimation on the DP images, where the defocus deblurring was guided via the depth estimation results. By introducing the assistance information of depth, better performance can be achieved. Nevertheless, these methods are all based on a single network to handle different regions in the DP image, while ignoring that the deblurring is spatially-varying for the DP data. To address this issue, our method focuses on adopting networks with different parameters to handle regions with different blur amounts, respectively. In this way, the information of the clear region can be well maintained (with a lighter network) and the missing details of the blurred region are prone to being recovered (with a heavier network).

\begin{figure*}[htbp]
    \centering
    \includegraphics[width=17cm]{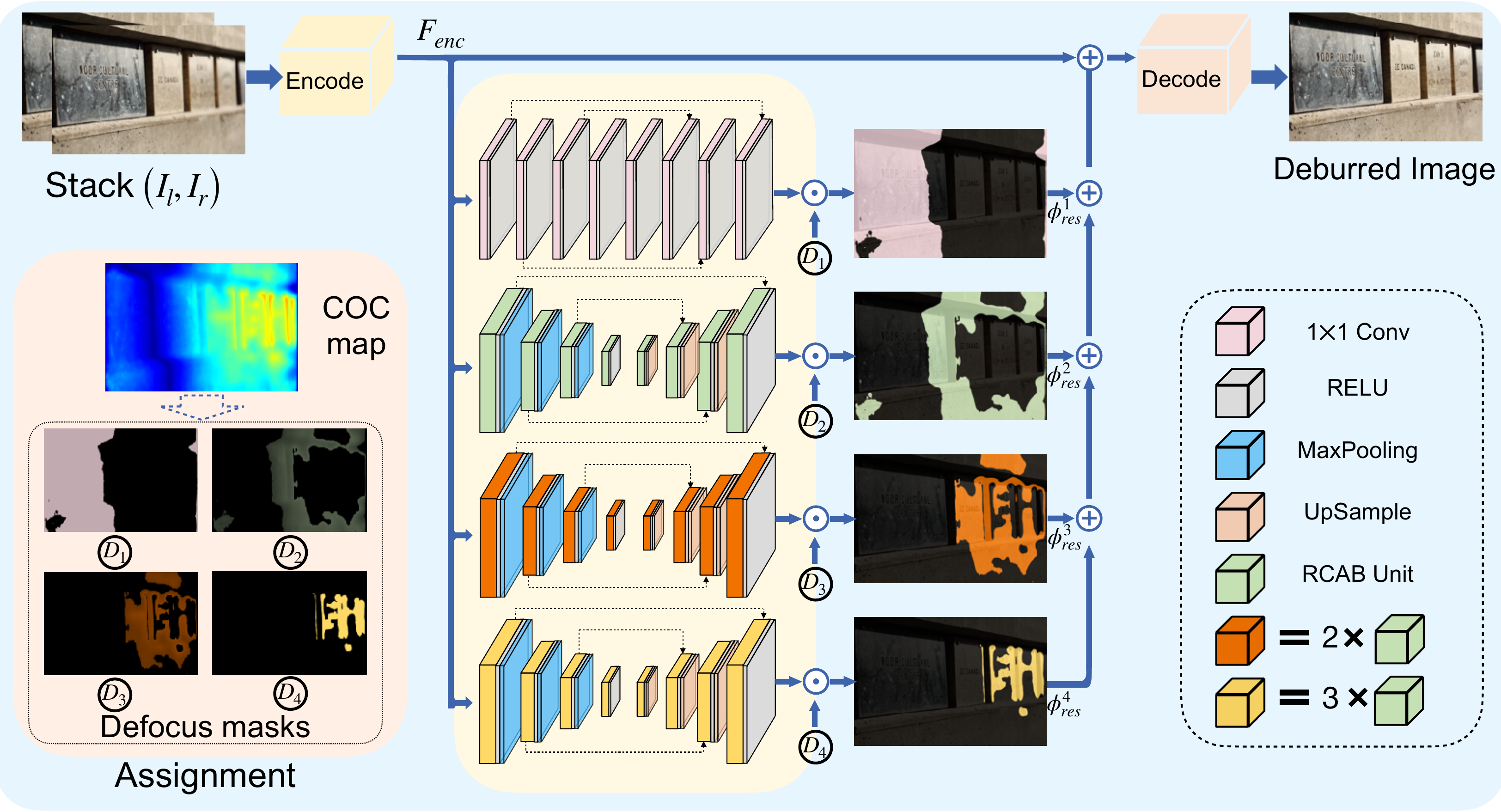}
    \caption{The workflow of our proposed BaMBNet. We stack the left view $I_l$ and right view $I_r$ and feed them into the encode module to extract the basic features. The COC map is used to generate the defocus masks, which can guide the multi-branch network to extract diversifiable residual features from the regions with different blur amounts.}
    \label{fig:framework}
\end{figure*}

\section{The Proposed BaMBNet Method}
\label{sec:Proposed}
From the definition of COC size Eq. (\ref{Imaging}), given the focal length $f_0$, focus setting $N$ and the subject-to-camera distance $d$, the blur amounts only rely on the diameter of the aperture $d_f$. Therefore, defocus blur often occurs when a narrow aperture is applied to a scene with a wide depth range. Since the COC size of one pixel is related to the depth range in the scene, when the COC size is within the allowable range, \emph{i.e.} the scene is within the DoF, the projected image is clear. As the depth range of objects gradually growing, the diameter of COC would monotonically increase. In other words, different regions of the image will have different blur amounts, which is mainly determined by the subject-to-camera distance.

Following the above observations, we proposed a blur-aware multi-branch network (BaMBNet) that consists of multiple different branches with different parameters. We wish the image regions are assigned to different branches guided by the COC map. Specifically, we introduce an assignment strategy for in-the-wild image with non-uniform defocus blur. Here, the assignment strategy is a mapping from COC map with continuous blur amounts to defocus masks with several discrete states. Generally, we expect that the lighter branch with few parameters to maintain the source in-focus regions and the heavier branch with more parameters to recover the image regions with the larger COC size.

In the following, we will first present the details of the BaMBNet. Then, we introduce how to predict a COC map and automatically transform the COC map to the defocus mask.

\subsection{Blur-aware Multi-branch Network}
The workflow of the proposed multi-branch network is shown in Fig. \ref{fig:framework}. Our method takes 6-channel DP data $I_{con}$ as input, which is generated by stacking the right and left views (the two RGB images have a total of 6 channels).
The proposed BaMBNet consists of an encoder $\phi_{encode}$, multiple branch residual bottleneck modules $\Phi = \{\phi_i; i=1,...,M\}$, and a decoder $\phi_{decode}$.
In this framework, the design intention of the encoder and decoder is to simply transform features between the image space and the latent representation space. Moreover, multiple bottleneck modules are used to extract the residual features to reconstruct the details adaptively.

In this paper, we represent the basic features extracted by encoder as $F_{enc}$. Afterward, the bottleneck modules $\Phi$ take the $F_{enc}$ as input and output a group of residual features guided by the defocus masks $\{D_i;i=1,...,M\}$, as shown in Fig. \ref{fig:framework}. The defocus masks indicate the blur level of the image, which are computed via combining COC maps and the thresholds. We will give details about estimating COC maps and solving the optimal thresholds in Section \ref{coc estimation} and Section \ref{assignment}, respectively. Finally, the group of residual features $\{\phi_{res}^i; i=1,...,M\}$ will be summed up to obtain the global residual features, and the decoder takes the sum of the basic features $F_{enc}$ and global residual features as inputs to transform these features into the target image $\hat{I}$. In summary, we can formulate the process as follows:
\begin{gather}
     F_{enc}  = \phi_{encode}(I_{con}), \notag \\
    \hat{I} = \phi_{decode}\left( F_{enc} + \sum_{i=1}^{M} D_i \odot \phi_i \left(F_{enc}\right)  \right ).
\end{gather}

Note that the architecture of bottleneck modules consists of two main types: the FCN-like network (the lightest branch) and the U-net-like network (other branches) \cite{long2015fully,ronneberger2015u}. Both two types of networks are symmetric structures. The FCN-like network only consists of eight $1 \times 1$ convolution layers and activate layers, while the U-net-like network contains 8 blocks: 4 encode blocks and 4 decode blocks. We expect the FCN-like network to focus on the clear and slightly blurred image regions.
Since the FCN-like network is the lightest branch, it does not need a large receptive field to recover the image regions and its goal is to maintain the information of image regions and remove the some slight noise.
Compared with the FCN-like network, the U-net like network has a more complex structure by introducing some up-sampling layers and max-pooling layers. Therefore, it can better capture the multi-scale features of the DP images.
Furthermore, the U-net-like network employs the residual channel attention blocks (RCAB) module that is widely used as a basic unit in super-resolution task \cite{zhang2018image}.
To recover image regions with different blur amounts, the U-net-like networks (different branches) are required to hold different capacities, \emph{i.e.} parameters. In practice, we simply change the number of RCAB modules in each block to adapt the capacities of different branches.
From the lightest U-net-like bottleneck module to the heaviest one, the number of RCAB modules in each block increases from 1 to 3. In addition, the output of every bottleneck module has the same channel number with the basic feature $F_{enc}$ to allow the residual features $\phi_{res}^i$ to add with $F_{enc}$.

In the training phase, we use the $L1$ loss function between the output $\hat{I}$ and the ground truth $I_{gt}$:
\begin{equation}
    \mathcal{L}_{rec} = \frac{1}{n} \sum^n \left \| I_{gt} - \hat{I}\right\|_1,
    \label{reconstruct loss}
\end{equation}
where $n$ denotes the number of samples.


\subsection{The COC Estimation}
\label{coc estimation}
Different from the existing end-to-end deep networks \cite{pan2020dual, abuolaim2020defocus}, which apply a single network to handle different regions in the DP image, the proposed BaMBNet treats different regions (with different blur amounts) by different networks (with different capacities).
Therefore, how to estimate the blur amounts, which can be measured via a COC map, in different regions of image is a crucial step in our proposed method. In this section, we will introduce a method to estimate the COC map from the DP image pairs in an unsupervised way.

Firstly, according to the analysis in \cite{punnappurath2020modeling}, the blur kernel of right and left views should be symmetrical
\begin{equation}
    H_r = H_l^f,
    \label{ kernel property}
\end{equation}
where $H_r$ and $H_l$ denote the blur kernels of the right and left view, respectively. $H_l^f$ represents the left blur kernel flipped along the vertical axis.

Since the symmetry property of the kernels, based on the assumption that any image patch $G$ with a fixed size has a constant-depth, the following corollary can be derived \cite{harikumar1999perfect, punnappurath2020modeling}:
\begin{equation}
    G_l * H_r = G_r * H_r^f,
    \label{intersting loss}
\end{equation}
where $G_l$ and $G_r$ denote the left and right DP views of $G$, $*$ denotes the convolution operation, and it provides an unsupervised solution for estimating the COC.

In this paper, we break the limitation of fixed size in image patch $G$ and propose an extension version of Eq. (\ref{intersting loss}) as:
\begin{gather}
    I_l^g = I_r^g,  \notag \\
    I_l^g = \{ \mathcal{N}_j I_l * H_r(j; \lfloor \hat{c(d)} \rceil)   \}, \notag \\
    I_r^g = \{ \mathcal{N}_j I_r * H_r^f(j; \lfloor \hat{c(d)} \rceil) \},
    \label{simplified differential}
\end{gather}
where $\hat{c(d)}$ is the estimated COC size and $\lfloor \cdot \rceil$ denotes an $\mathrm{round}$ function to satisfy the physical constraint of the blur kernel. $\mathcal{N}_j I_l$ denotes the image patch of size $ \hat{c(d)} \times \hat{c(d)}$ extracted at location $j$, and $\mathcal{N}_j$ is the matrix extracting patch $\mathcal{N}_j I_l$ from $I_l$ at location $j$. The meaning of $\mathcal{N}_j I_r$ is similar. The $\{ \cdot \}$ denotes a concatenation operator which concatenate the pixels into a map of the same size as $I_l$ and $I_r$.
Here we postulate that the COC size is equal to the radius of a local neighborhood, in which the image patch corresponds to a constant-depth. Since the COC size is usually small enough, the assumption can be regarded as an application of Riemann integral, which is more reasonable than the fixed size of image patch in Eq. (\ref{intersting loss}). When determining the radius of the blur kernel, the weight of $H_l, H_r$ are also determined by referring to the translating disk proposed by Abhijith \emph{et. al} \cite{punnappurath2020modeling}.

\textbf{Loss Function.} According the Eq. (\ref{simplified differential}), we can intuitively formulate the loss function as:
\begin{equation}
    \mathcal{L}_{gem} = \| I_l^g - I_r^g \|.
    \label{unsurprivise loss}
\end{equation}

However, the loss (\ref{unsurprivise loss}) does not work well because the in-the-wild images are usually affected by noise. To relieve the problem, we apply Gaussian blur to smooth the residual results before the $l_2$ norm.
Furthermore, we add a prior regularization term for the unsupervised geometric loss $\mathcal{L}_{gem}$, which penalizes the gradient of the network output and smoothes the estimated COC map. Finally, the total loss of COC estimation can be formulated as:
\begin{equation}
    \mathcal{L}_{coc} = \mathcal{L}_{gem} + \lambda \left \| \nabla(I_{COC}) \right \|,
\end{equation}
where $I_{COC}$ denotes the estimated COC map, and $\lambda$ is a balance factor for the geometric loss term and the regularization term. In the COC map estimation task, we use almost the same network architecture as the multi-branch network described in Fig. \ref{fig:framework} without defocus masks. The only difference is that we remove the last activation layer to allow the network to output both negative and positive values. The sign of radius indicates the relative positional relationship between the scene and the focus plane.

\subsection{Meta-learning Defocus Mask Generation}
\label{assignment}
Given the COC map, by assigning each pixel to a proper branch according to the blur amounts (COC values), we expect that clear regions can be preserved with lighter branches and blurred regions can be well recovered with heavier branches. In this section, we will introduce a defocus mask generation method to divide the continue COC map into a limited number of levels, where each level corresponds to a defocus mask (as shown in Fig. \ref{fig:framework}).

An intuitive strategy is to divide the COC size into different levels by some pre-defined thresholds. However, humancrafted thresholds are sub-optimal and not suitable for all images. In the following, we will present an optimization method that jointly optimizes both thresholds and multi-branch network parameters. Here, the thresholds can be seen as the hyper-parameters of the network, and they are very difficult to be determined. In this paper, we introduce a method capable of adaptively learning the thresholds directly from a small amount of meta-data (e.g., the validation dataset), thus they can be finely updated simultaneously with the learning process of the network parameters.

In particular, the defocus mask generation can be transformed into finding some thresholds $\{r_1,...,r_{M+1}\}$ of one image and assign each pixel $p$ at position $(p_x,p_y)$ to the corresponding mask by $q(p;r)=i$, where $i$ is the mask index of pixel $p$. Here, $r_1$ and $r_{M+1}$ are the predefined minimum and maximum COC value, respectively. Further, when the COC size $r$ of pixels is in the range of $[r_i,r_{i+1}]$, we assign these pixels to the branch bottleneck $\phi_i$. $r_i$ and $r_{i+1}$ are the lower and upper bounds, respectively.



To measure the assignment errors, we build a \emph{validation dataset} to measure the disparity between the deblurring result $\phi_i(p;r)$ and its ground truth $I(p)$.
The total loss $\mathcal{L}_{val}$ can be formulated as:
\begin{equation}
    \mathop{\arg \min}\limits_{\{r_1,...,r_{M+1}\}} \sum_{i=1}^{M}  \int_{r_i}^{r_{i+1}} \left| \phi_i(p;r) - I(p) \right| \mathrm{d}r.
    \label{Quantization loss function}
\end{equation}

Combined with the model training process, generating the defocus mask is a nested optimization: the inner optimization trains multi-branch networks $\Phi$ given thresholds $\{r_1,...,r_{M+1}\}$. The outer optimization will evaluate the trained network $\Phi$ and fine-tune the thresholds $\{r_1,...,r_{M+1}\}$ by computing the assignment errors on the validation dataset:
{
\begin{equation}
    \mathop{ \arg \min}_{\{r_1,...,r_{M+1}\}} \mathcal{L}_{val} \left (  \mathop{ \arg \min}_{\Phi}\mathcal{L}_{rec}(\Phi,\{r_1,...,r_{M+1}\}) \right).
    \label{Iteration loss}
\end{equation}
}

In actually, our iteration solution is inspired by the hyperparameters optimization in the meta-learning schemes \cite{lorraine2018stochastic}. The differences are that we adopt a gradient-free optimization method due to the gradient of threshold $r_i$ is unavailable. The details of the algorithm are shown in Alg. \ref{quatization algorithm}.


\begin{algorithm}
\DontPrintSemicolon
    \SetKwInOut{KwInput}{Input}
    \SetKwInOut{KwOutput}{Output}
    \KwInput{The number of level $M$, multi-branch bottleneck modules $\Phi = \{\phi_i; i=1,...,M \}$}
    \KwOutput{The optimal thresholds $\{r_i; i= 1,...,M+1 \}$}
    Initialize the thresholds uniformly. Set $M$ to 4, and fix the minimum threshold to $r_1=0$ and the maximum threshold to $r_5=25$. \;
    \While{do not converge}{
        Given the $r_2, r_3, r_4$, optimizing the multi-branch bottleneck modules $\Phi={\phi_1, \phi_2, \phi_3, \phi_4}$ on the training dataset until the model is convergence.\;
        With the reference to loss (\ref{Quantization loss function}), compute the costs of multi-branch bottleneck modules $\Phi$ on each available COC size $r$ on the \emph{validation dataset}, and assign the COC size $r$ of pixel to the proper branch. Then update the thresholds $r_2, r_3, r_4$ according to (\ref{Iteration loss}).
    }
\caption{Meta-learning Defocus Mask Generation Algorithm}
\label{quatization algorithm}
\end{algorithm}

\subsection{Implemental Details}
We train our proposed BaMBNet in two steps: 1) training the COC estimation network to obtain the COC map, 2) joint training of the defocus deblurring network and determining the $M+1$ thresholds.

The COC map prediction is an unsupervised task. In our experiments, we directly use almost the same network architecture as the multi-branch network described in Fig. \ref{fig:framework}. Here we remove the point-wise multiplications with defocus masks and also remove the last activation layer to allow the network to output both negative and positive values.
During estimating the COC map, we set the hyper-parameter $\lambda$ to 10 and run 10 epochs at the learning rate of $2e$-5. Referring to \cite{punnappurath2020modeling}, we set 25 as the upper bound of estimated COC size.

Then, when we determine the thresholds through the assignment strategy, the number of bottleneck modules $M$ is set to 4, the minimum $r_1$ and maximum $r_{M+1}$ thresholds are set to 0 and 25, respectively. We uniformly assign to the initial thresholds $r_2,r_3,r_4$ and update them at every 5 epochs until the iteration converges (at around 45 epochs). The initial learning rate is set to 2$e$-4. And then, we fix the thresholds and try to train the multi-branch network. The initial learning rate starts from 2$e$-4 which is decreased by half every 60 epochs. In addition, we use an annealing strategy to train our end-to-end network. Specifically, in the early training phase, the generated output $\hat{I}$ will rely on the guidance of the defocus masks $D_1,D_2,D_3,D_4$.
As the model will gradually converge, we gradually reduce the weights associated with the masks until they are zeros. The annealing strategy will bring two advantages.
First, the network can adaptively optimize the entire images to avoid discontinuous edges due to stitching residual features from different branches.
Second, our network no longer needs the defocus masks in the testing phase, which facilitates the deployment of our method. We apply the Adam optimizer with mini-batches of size 2 \cite{die2015adam} to optimize our model.
Our model is trained on a computer equipped with an AMD 2.0GHz CPU, 32G memory, and an RTX 3090 GPU, and converges after 200 epochs. We implement our method in the PyTorch framework.

\section{Experiments and Results}
\label{sec:Experiments}

In this section, we evaluate our proposed BaMBNet model on the defocus deblurring task. Firstly, we compare our method with four state-of-the-art methods on the DPD-blur dataset \cite{abuolaim2020defocus}. Then, we make thorough ablation studies to demonstrate the effectiveness of the proposed method.

\subsection{Evaluation on DPD-Blur Dataset}
We pre-processing the training dataset following the settings of DPDNet totally \cite{abuolaim2020defocus}. Specifically, a slide window of 512$\times$512 pixels is applied to crop image patches on the training image of 1680 $\times$ 1120 with 60\% overlap. By computing the sharpness energy, we discard 30\% the most homogeneous regions in the crop patches. 

To verify the effectiveness of the proposed method, we compare four methods such as the edge-based defocus blur (EBDB) \cite{karaali2017edge}, the defocus map estimation network (DMENet) \cite{lee2019deep}, the dual-pixel defocus deblurring network (DPDNet) \cite{abuolaim2020defocus}, and the DP-based depth and deblur network (DDDNet) \cite{pan2020dual}.

Note that EBDB \cite{karaali2017edge} and DMENet \cite{lee2019deep} are proposed for defocus map estimation, and cannot be directly applied for defocus deblurring. Followed the advice by \cite{abuolaim2020defocus}, we additional leverage a non-blind deblurring method with the defocus map \cite{krishnan2009fast} to recover the defocus blur images. Since DPDNet \cite{abuolaim2020defocus} shares the same experimental settings with our method, we directly evaluate the trained model they has already provided for best performance. DDDNet \cite{pan2020dual} is the latest method and the code is not available now, so we re-implement this method and train their model with our training samples.

\textbf{Evaluation metrics.} All methods are evaluated by five metrics: peak signal-to-noise ratio(PSNR), structural similarity(SSIM) \cite{wang2004image}, mean absolute error(MAE), learned perceptual image patch similarity(LPIPS) \cite{zhang2018unreasonable} and Fr\'{e}chet Inception Distance (FID) \cite{heusel2017gans}. The PSNR, SSIM and MAE provide a traditional standard measurements in reconstruction errors, while the LPIPS and FID supplement a similarity judgment from the human and semantic perceptions.

\subsubsection{Quantitative Results}

\begin{table*}[htbp]
\setlength\tabcolsep{8pt}
\centering
\footnotesize
\caption{The quantitative evaluation results in terms of PSNR, SSIM, and MAE for different defocus deblurring methods on the DPD-blur dataset. The bold numbers indicate the best results while the second bests are marked by underlines.}
\label{tab:quantitative comparison metrics}
\resizebox{0.95\textwidth}{!}{%
\begin{tabu}{c|c|c|c|c|c|c|c|c|c|c}
\toprule[0.75pt]
\multicolumn{1}{c|}{\multirow{2}{*}{Method}}  & \multicolumn{3}{c|}{Indoor} & \multicolumn{3}{c|}{Outdoor} & \multicolumn{3}{c|}{Combined} & {\multirow{2}{*}{PARA(M)$\downarrow$}} \\ \cline{2-10}
\multicolumn{1}{c|}{} & PSNR$\uparrow$  & SSIM$\uparrow$  & MAE$\downarrow$  & PSNR$\uparrow$  & SSIM$\uparrow$  & MAE$\downarrow$ & PSNR$\uparrow$  & SSIM$\uparrow$  & MAE$\downarrow$ \\
\tabucline[0.75pt]{-} %
EBDB \cite{karaali2017edge}  & 25.77 & 0.772 & 0.040 & 21.25 & 0.599 & 0.058 & 23.45 & 0.683 & 0.049  & - \\ 
DMENet \cite{lee2019deep} & 25.50 &0.788 & 0.038 & 21.43 & 0.644 & 0.063 & 23.41 & 0.714 & 0.051  & 26.94 \\ 
DPDNet \cite{abuolaim2020defocus} & 27.48 & \underline{0.849} & \underline{0.029} & 22.90 & 0.726 & 0.052 & 25.13 & \underline{0.786} & \underline{0.041}  & 29.29 \\ 
DDDNet \cite{pan2020dual} & \underline{27.57} & 0.833 & 0.030 & \underline{23.28} & 0.708 & \underline{0.050} & \underline{25.36} & 0.768 & \underline{0.041}  & \underline{5.62} \\ 
Ours & \textbf{28.60} & \textbf{0.872} & \textbf{0.026}  & \textbf{24.30} & \textbf{0.772} & \textbf{0.045}  & \textbf{26.40} & \textbf{0.821} & \textbf{0.036} & \textbf{4.50} \\
\tabucline[0.75pt]{-} %
\end{tabu}%
}
\end{table*}

The defocusing deblurring results of different methods are reported in Table \ref{tab:quantitative comparison metrics}. We can learn that our method achieves significant improvements compared with other SOTAs.
For example, while DPDNet and our method use exactly the same training dataset, our method improves the performance of 1.27dB in term of PSNR and the parameters of our method are also reduced by 85\%. By introducing the depth information of the scenes, DDDNet \cite{pan2020dual} achieves the second best performance in most cases. The gain over these two SOTAs is mainly because that our multiple branches can cope with larger blur variations while keeping the in-focus region details.

\begin{table}[htbp]
\setlength\tabcolsep{5pt}
\centering
\footnotesize
\caption{The quantitative results in terms of LPIPS and FID for different defocus deblurring methods in the DPD-blur dataset.}
\label{tab:subjective comparison metrics}
\resizebox{0.485\textwidth}{!}{%
\begin{tabu}{c|c|c|c|c|c|c}
\toprule[0.75pt]
\multicolumn{1}{c|}{\multirow{2}{*}{Method}} & \multicolumn{2}{c|}{Indoor} & \multicolumn{2}{c|}{Outdoor} & \multicolumn{2}{c}{Combined} \\ \cline{2-7}
\multicolumn{1}{c|}{} & LPIPS$\downarrow$  & FID$\downarrow$  & LPIPS$\downarrow$  & FID$\downarrow$ & LPIPS$\downarrow$  & FID$\downarrow$ \\
\tabucline[0.75pt]{-} %
EBDB \cite{karaali2017edge} & 0.297 & 127.17  & 0.373 & 132.41  & 0.336 & 119.87 \\ 
DMENet \cite{lee2019deep} & 0.298 & 114.26 & 0.397 & 117.46 & 0.349 & 107.39 \\ 
DPDNet \cite{abuolaim2020defocus} & 0.189 & \underline{58.98} & 0.255 & \underline{53.12}  & 0.223  & \underline{52.05} \\ 
DDDNet \cite{pan2020dual} & \underline{0.158} & 65.01 & \underline{0.233} & 66.97 & \underline{0.197} & 61.17 \\ 
Ours   & \textbf{0.124}  & \textbf{46.75} & \textbf{0.157} & \textbf{39.92} & \textbf{0.141}  & \textbf{40.28} \\
\tabucline[0.75pt]{-} %
\end{tabu}%
}
\end{table}

Aside from the traditional metrics that are mainly used to measure the reconstruction errors, we also provide two recent perceptive metrics, LPIPS and FID, which are widely used to evaluate the perceptive quality of generated images on the low-level vision tasks. Rather than classic per-pixel measurements, perceptive metrics leverage deep semantic representations considering context-dependent and high-order image structures, which are more closed to human judgments of similarity. As shown in Table \ref{tab:subjective comparison metrics}, compared to the results of DPDNet, our LPIPS and FID indexes have decreased by 36.8\% and 22.6\%, respectively. It is well known that the LPIPS appears to be more sensitive to blur and FID reflects the similarity of images in high-dimensional space.
So that the higher performance perceptive metrics indicate that our method can generate more sharp images. When compared with DDDNet \cite{pan2020dual}, which is the latest and most competitive method, the proposed method achieves a considerable improvement in all metrics.

\subsubsection{Qualitative Results}

\begin{figure*}[htbp]
    \centering
    \includegraphics[width=18cm]{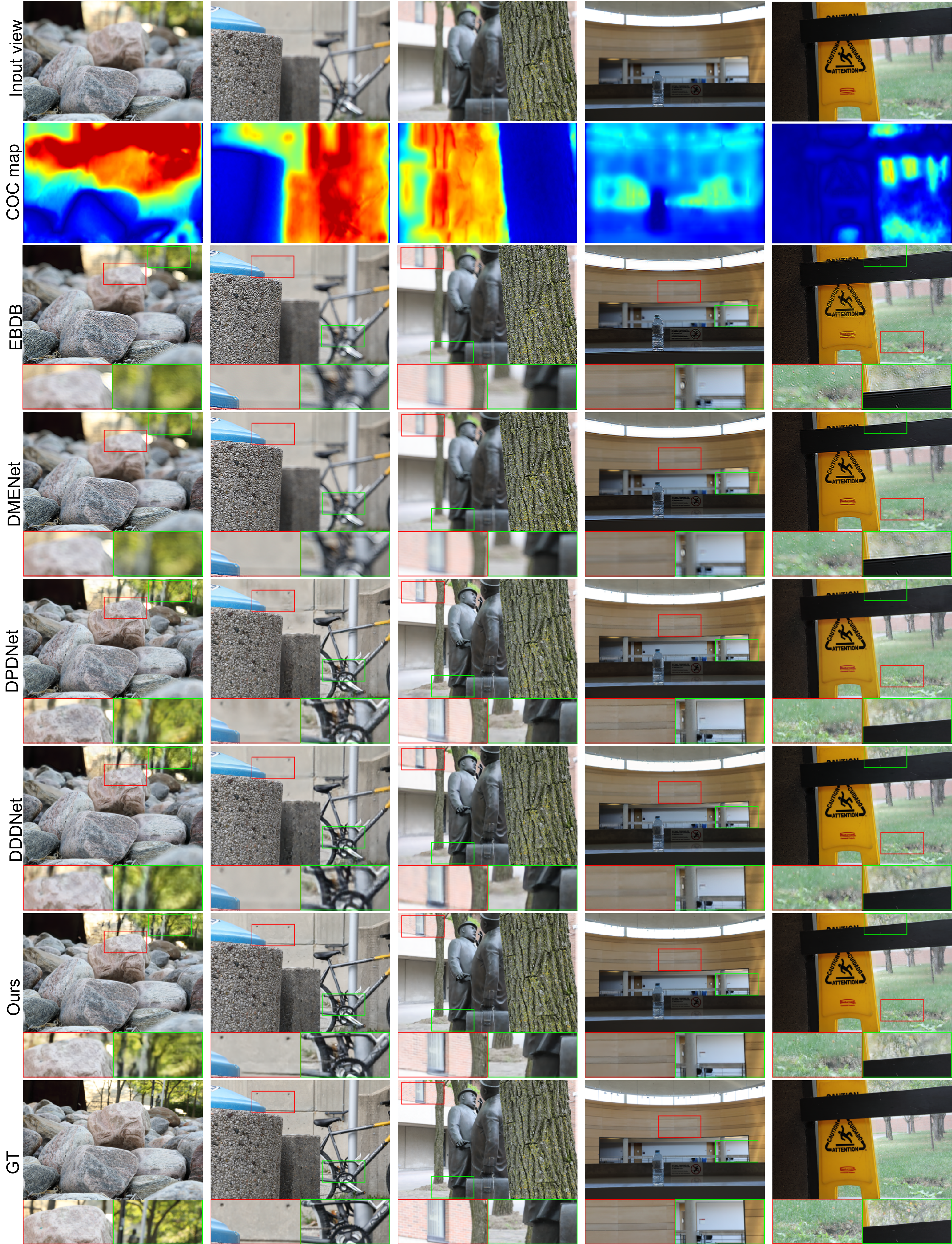}
    \caption{Qualitative results of the state-of-the-art deblurring methods. The second row is the estimated COC maps by our unsupervised method given the inputs of first row. We highlight the cropped patches by green and red boxes.}
    \label{fig:visual comparison.}
\end{figure*}

In Fig. \ref{fig:visual comparison.}, we present defocus deblurring results by different methods including five representative scenes, \emph{i.e.} the 1st-3rd columns show the outdoor scenes, 4-5th columns show the indoor scenes. It demonstrates that our deburred method is robust to varying scenes and our results have better visual quality.
For example, the DPDNet \cite{abuolaim2020defocus} not only generates unexpected artifacts on the region with a wider depth, but also fail to remove the defocus on the region with the widest depth, where are highlighted by red and green boxes in the first column, respectively. While our method can appropriately reconstruct the blur region and well maintain the sharp region, as shown by highlighted zoomed-in boxes. These results demonstrate that our BaMBNet can simultaneously cope with multiple level blur amounts on an image with a wide depth range. For another example, in the fourth column, most deblurring methods can keep the information very well in the in-focus region. However, as the defocus blur distortion becomes increasingly serious, they may all tend to generate smooth deblurring results and miss the details. Particularly, the result of DDDNet misses the stride textures in the highlighted region in Fig. \ref{fig:visual comparison.}. Compared with DDDNet \cite{pan2020dual}, our method is able to produce more sharp results.

From the comparison results, we also learn that both DPDNet \cite{abuolaim2020defocus} and DDDNet \cite{pan2020dual} will suffer from performance degradation when encountering hard samples. As shown in the fifth column of Fig. \ref{fig:visual comparison.}, the highlighted region shows rich details on the window, while the background exhibits seriously blur. However, both DPDNet \cite{abuolaim2020defocus} and DDDNet \cite{pan2020dual} fail to deal with these cases. They cannot well preserve the detail of windows (is slightly blurred when compared with the ground truth) and remove the blur of background. On the contrary, our method successfully achieves a better balance between preserving windows detail and deblurring the background. This indicates that our network is more efficient to handle complex scenes compared with other methods.

\subsection{Ablation Studies}
\begin{table}[t]
\centering
\footnotesize
\caption{Results of defocusing deblurring with or without the defocus masks on the DPD-blur dataset. }
\label{tab:without quantization}
\resizebox{0.45\textwidth}{!}{%
\begin{tabu}{l|l|l|l|l|l}
\toprule[0.75pt]
Models & PSNR$\uparrow$  & SSIM$\uparrow$  & MAE$\downarrow$   & LPIPS$\downarrow$  & FID$\downarrow$\\
\tabucline[1pt]{-} %
w/o-defocusmask & 26.22 & 0.815 & 0.036  & 0.151 & 42.34\\ 
w-defocusmask  & \textbf{26.40} & \textbf{0.821} & \textbf{0.036}  & \textbf{0.141}  & \textbf{40.28}\\ 
\tabucline[0.75pt]{-} %
\end{tabu}%
}
\end{table}

To demonstrate the effectiveness of the proposed meta-learning defocus masks generation, we report the objective results of our method with and without the defocus masks in Tab. \ref{tab:without quantization}. Indeed, compared to directly training the network without distinguishing the blur amounts, leveraging defocus masks to assign the pixels to different branches shows a considerable gain (0.2 dB higher in PSNR). In addition, training with the assistance of defocus masks is able to obtain more sharp results, which can be verified by the improvement in term of LPIPS.

\begin{figure*}[htbp]
    \centering
    \includegraphics[width=17.2cm]{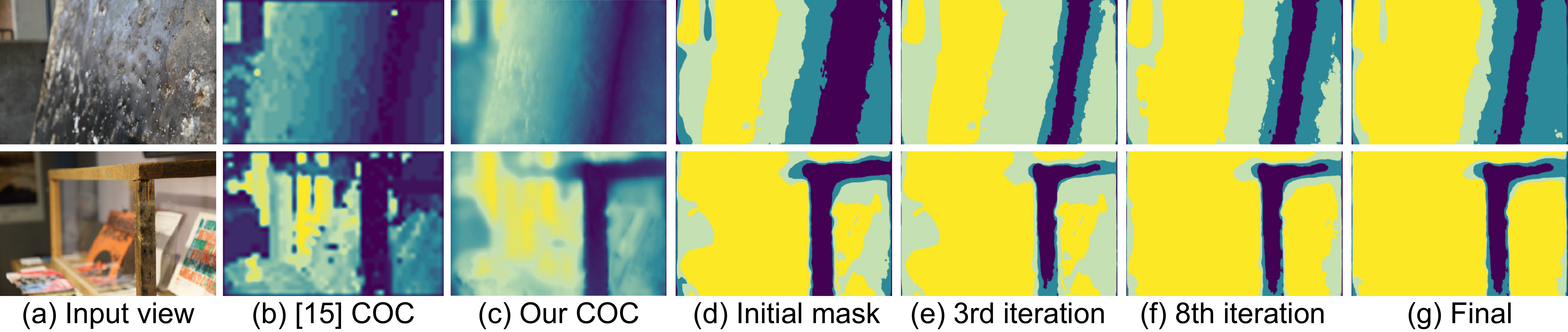}
    \caption{Comparisons between defocus masks generated in different iteration optimization epochs, and the qualitative comparison results of COC map. (b) and (c) show the COC maps estimated by \cite{punnappurath2020modeling} and our end-to-end model. Yellow, LightBLue, Blue, and Purple represent Heavy, Medium, Light, and Slight blur amounts respectively. }
    \label{fig:iteration process}
\end{figure*}

\begin{figure*}[htbp]
    \centering
    \includegraphics[width=0.95\linewidth]{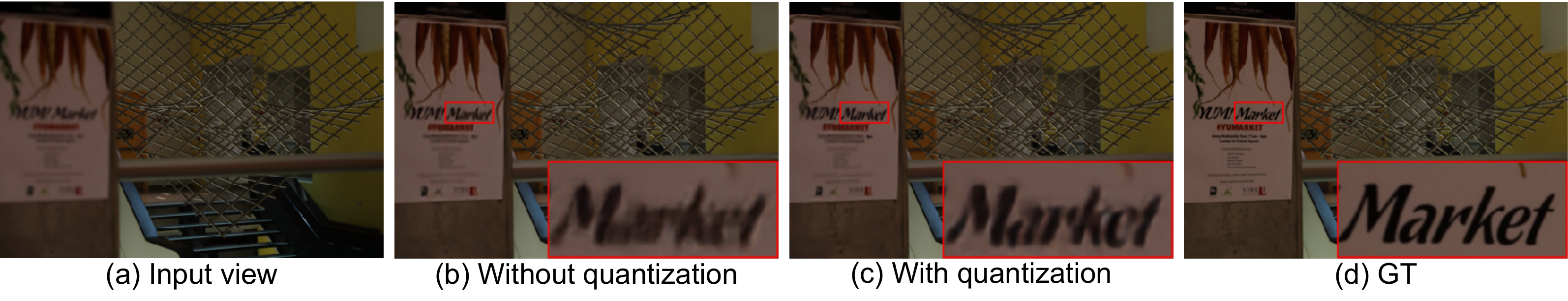}
    \caption{Qualitative comparisons with and without the assistance of defocus masks. (a) shows the input view. (b) shows the result of without defocus masks. (c) shows the result of our normal network. We also display the zoom-in cropped patches for comparison. Since (b) and (c) share totally the same network architecture, (c) shows better deblurring performance than (b) under the guidance of defocus masks. (d) shows the corresponding ground truth.}
    \label{fig:quation}
\end{figure*}

Apart from the quantitative comparisons, we also show the generated defocus masks in different iteration epochs. As shown in Fig. \ref{fig:iteration process}, the initial defocus masks split the entire image into four separate groups (here we directly divide the entire interval of COC into four equal parts). After several iterations, the masks are adjusted to be more reasonable. For example, the regions marked by yellow are correspond to the image regions with large blur amounts. Besides, the defocus masks do not require pixel-wise preciseness because the masks work on deep features that represent semantic information. Generally, the regions with more blur amounts usually tend to use the heavier branch for restoration, while the slightly blurred regions are assigned to the lighter branches with a small receptive field to avoid the artifacts. These results indicate that the refined defocus masks are able to offer effective guidance information (\emph{i.e.} the blur distribution of image) for multi-branch bottleneck modules.

\begin{figure}[hp]
    \centering
    \includegraphics[width=1\linewidth]{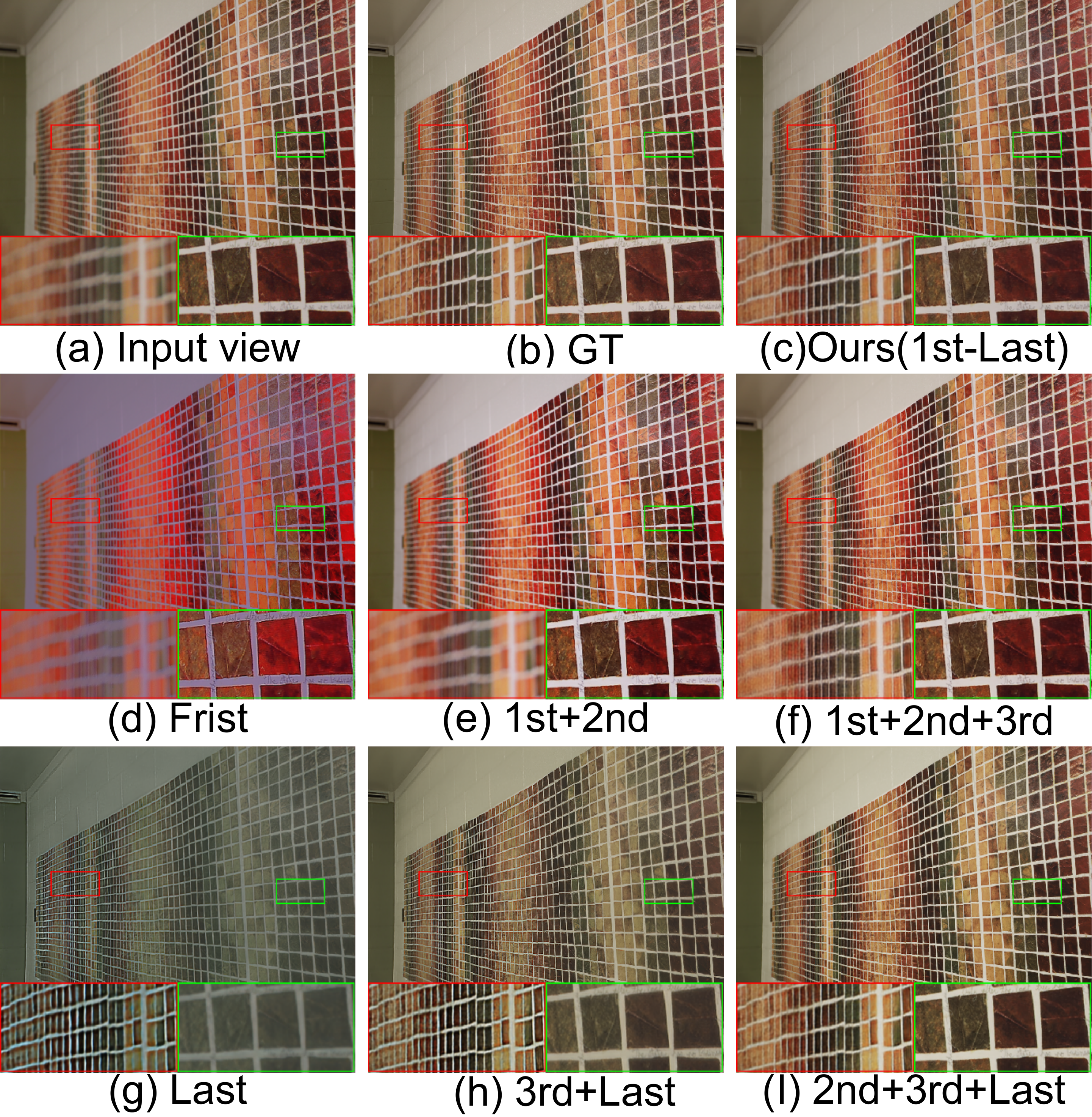}
    \caption{Quantitative results of network with variant residual branch bottlenecks. (a) and (b) show the input view and ground truth, respectively. (c) shows the results of our network. We also try to investigate the role of each branch in the entire network by removing the residual features of some branches. In the second row, the comparison models first harness the residual features of the lightest branch in the 1st column and add more residual features in heavier branch in 2nd and 3rd columns. In the third row, we first employ the comparison model with the heaviest branch in the 1st column and add lighter branch in the next two columns.}
    \label{fig:level}
\end{figure}

The qualitative results with and without defocus masks are shown in Fig. \ref{fig:quation}, the networks trained directly by removing the defocus masks fail to recover the word, \emph{i.e. `market'}, in which these deblurred characters are difficult to recognize. It is because that the network trained without the assistance of defocus masks would treat the blurred characters and other clean regions equally, and it is difficult for the network to focus on the deblurring of blurred characters. As for the network trained with the assistance of defocus masks, it will easily distinguish the difference between blurred characters and other clear regions, thus focusing on these blurred regions (by assigning a heavy network large capacity to these regions).
In summary, our network achieves performance improvement both in objective metrics and visual results with the assistance of defocus masks.

\begin{figure*}[htbp]
    \centering
    \includegraphics[width=17cm]{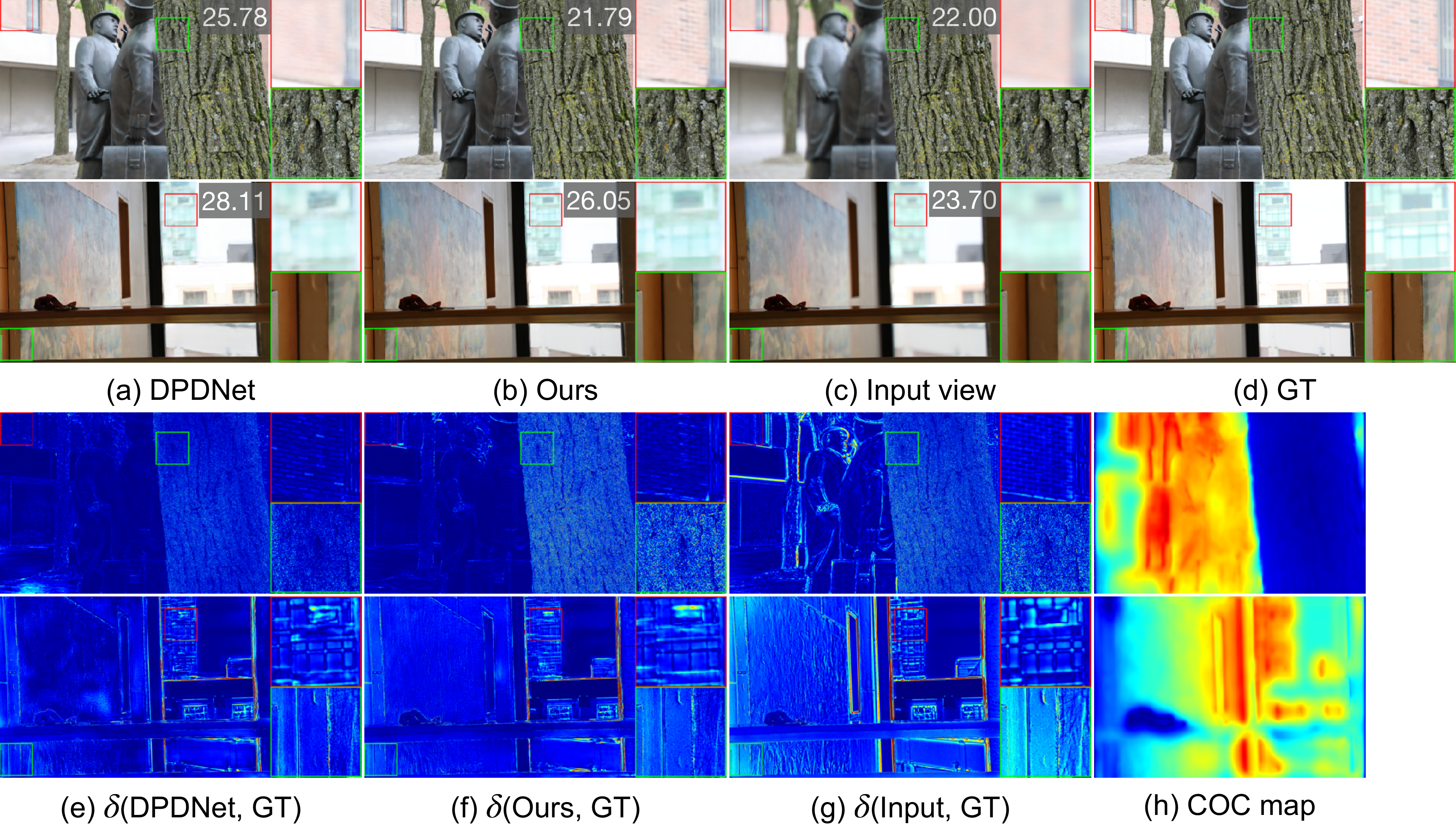}
    \caption{Example failure measure cases of our method. The number shown on images in the first two rows is the PSNR measures compared with GT. $\delta$ is enhanced residual map of given images.}
    \label{fig:failure analysis}
    \vspace{-0.5cm}
\end{figure*}


We adopt an extremely strict way to explore the role of the residual features extracted from each branch bottleneck. Fig. \ref{fig:level} shows the quantitative results of comparison models. Instead of retraining the network with a different number of branch for comparison, the comparison models directly drop the residual features of some branches from the pretrained model. Just as one expected, when the lighter branch with less output residual features takes part in deblurring, the network pays attention to preserve the details in the in-focus region. For instance, the in-focus region highlighted by the green box keeps a lot of texture information as shown in Fig. \ref{fig:level}(d). As more residual features of the heavier branches gradually participate into the deblurring, the defocused blur regions become gradually sharper and begin to recover more texture details from the blurred regions. Aside from exploring the effect of the lightest branch, we also present the results of the comparison model with only the heaviest branch in Fig. \ref{fig:level}(g). Compared with the result in Fig. \ref{fig:level}(d), the blur region of Fig. \ref{fig:level}(g) shows remarkable performance which is highlighted by the red box, while the in-focus region in the green box loses detailed information. It successfully demonstrates that the heavier residual branch just focuses on the larger blur region. Our method not only assigns the blurred regions into a group of residual branch bottlenecks according to the blur amounts, but also achieves the task separation of preserving the details in the focus region and deblurring in the blur region.

\subsection{Some Failure Cases in Term of PSNR}
Overall, the performance of our proposed BaMBNet is much better than the SOTA methods when adopting the PSNR metric. However, when traversing the PSNR result in the test dataset, we find that some of our samples show a considerable decrease gap when compared with the SOTAs. To investigate the reasons, we present the two worst cases (\emph{e.g.}, 2 dB decreases in term of PSNR) and provide the zoomed-in patches of the most blur and clear regions in Fig. \ref{fig:failure analysis}. Although our method achieves lower PSNR results, it dramatically alleviates the blur on the most blurred regions as shown in the red boxes. Moreover, our results do not show visible differences in the clear regions highlighted by the green boxes. Therefore, it can be proved that our method tends to obtain clear and sharp results even if the PSNR result is poor. To explain the difference between PSNR measurement and human perceptions, we present the residual images between the results and ground truths in the last two rows of Fig. \ref{fig:failure analysis}. We found an interesting phenomenon: the values of the residual images between the input images and ground truth are not zeros at the focus region. These values (which should be zeros) are even larger than these between DPDNet method and the ground truth. Therefore, we guess that both the ground truth and the input images have been disturbed by noise or they are misalignment.

\section{Conclusions}
\label{sec:Conclusions}
In this paper, we propose a blur-aware multi-branch network (BaMBNet) to deblur the real defocused images with nonuniform blur amounts. The BaMBNet automatically applies the lighter branch to the clear regions to maintain the detailed information, while assigning the heavier branch to the blurred region to recover the latent structural edge. In particular, we first devise an unsupervised scheme to estimate the COC map by the DP data, then generate defocus masks from the COC map by an assignment strategy. The defocus masks indicate the mapping from regions to branches. At last, we employ the defocus masks to guide the multi-branch bottlenecks with different parameters to handle the blur regions with different recovered complexity. Therefore, the overall optimal complex problem is divided into multiple simple subproblems, each branch only needs to solve the assigned subproblems, which allow to avoid the overfitting on the clear region and under-fitting on the most blurred region. Experimental results show that our BaMBNet performs robust to defocus blur in a wide depth range and achieves better balance between maintaining the information of clear regions and generating sharp details for the blurred regions.

\ifCLASSOPTIONcaptionsoff
  \newpage
\fi



%
{
\bibliographystyle{IEEEtran}
\bibliography{BaMBNet}
}

\end{document}